\documentclass[showpacs,preprintnumbers,twocolumn,prd]{revtex4}
\usepackage{amsmath,amssymb}
\usepackage[dvips]{graphicx}

\newcommand{\MeV}{\;\text{MeV}}

\begin{document}
\title{Critical surface in hot and dense QCD
       with the vector interaction}
\author{Kenji Fukushima}
\affiliation{Yukawa Institute for Theoretical Physics,
 Kyoto University, Kyoto 606-8502, Japan}
\begin{abstract}
 We discuss the chiral phase transition of hot and dense quark
 matter.  We illustrate that the first-order phase transition is
 generally favored at high baryon density and the repulsive
 vector-vector interaction weakens the first-order phase transition.
 We use the Nambu--Jona-Lasinio model with the Polyakov loop coupling
 for concreteness.  We locate the QCD critical surface on the quark
 mass plane for various values of the vector coupling constant.  We
 find that, with increasing quark chemical potential, the first-order
 region in the quark mass plane could shrink for sufficiently large
 vector coupling.  This may be a possible explanation for the recent
 lattice QCD results by de~Forcrand and Philipsen.
\end{abstract}
\preprint{YITP-08-76}
\maketitle


\section{Introduction}

On the phase diagram in the plane with the axes by temperature,
density, pressure, concentration, external fields, etc, the ``critical
point'' commonly refers to the terminal point of the phase transition
boundary of first order.  This point has an exact second-order phase
transition regardless of symmetry properties.

Interestingly enough, the existence of the critical point is a
possibility not only in condensed matter physics but also in QCD
(Quantum Chromodynamics) physics~\cite{Stephanov:1998dy}.  The QCD
phase diagram with the temperature $T$ and the quark chemical
potential $\mu$ has taken on significance in the application to the
heavy-ion collisions at various energies as well as to the neutron
star structure.  There seems to be a consensus that the QCD phase
transitions associated with chiral symmetry restoration and color
deconfinement are smooth crossover when $T$ goes up with $\mu\simeq0$.
In contrast, the situation changes in the different regime where $\mu$
grows with $T\simeq0$.  Model
studies~\cite{Asakawa:1989bq,Halasz:1998qr} suggest that the
first-order chiral phase transition should occur at some $\mu$
comparable to the constituent quark mass (or one third of the baryon
mass) when $T\simeq0$.  To be consistent with crossover at $T\neq0$
and $\mu\simeq0$, therefore, the critical point is expected to exist
at intermediate $T_E$ and $\mu_E$.

It is, however, important to remark that the existence of the QCD
critical point is still under dispute~\cite{Fukushima:2008pe}.  One
interesting negative observation against the QCD critical
point has come from the Monte-Carlo simulation of QCD on the lattice
with varying the light-quark mass $m_{ud}$ and the strange-quark mass
$m_s$~\cite{deForcrand:2006pv}, though some other lattice simulations
are rather affirmative.  (For a review see
Ref.~\cite{Schmidt:2006us}).

For the symmetry reason the chiral phase transition in three-flavor
quark matter at $m_{ud}=m_s=0$ is presumably first
order~\cite{Pisarski:1983ms}.  Because the mass term explicitly breaks
chiral symmetry, the first-order transition turns to crossover at some
$m_{ud}$ and $m_s$~\cite{Brown:1990ev,Gavin:1993yk}, which defines the
critical boundary in the $m_{ud}$-$m_s$ plane.  Model studies that
support the QCD critical point predict that the first-order region in
the $m_{ud}$-$m_s$ plane expands by the effect of increasing $\mu$, so
that the physical quark mass point hits the critical surface at
$\mu=\mu_E$~\cite{Kovacs:2006ym,Fukushima:2008wg}.  However,
de~Forcrand and Philipsen~\cite{deForcrand:2006pv} recently claim that
the first-order region should not expand but rather shrink at higher
density as long as $\mu/T$ is small.

The purpose of this paper is twofold:

1) We will extract the general feature of the quasi-particle
description at high density to favor the first-order phase transition.
We make use of simple closed expressions by limiting our discussions
to the $T=0$ case.  In fact, the confirmation of the first-order phase
transition at $T=0$ and $\mu\neq0$ suffices for the existence of the
QCD critical point since the QCD phase transition at $\mu=0$ and
$T\neq0$ is crossover.  In the same way we discuss the effect of the
repulsive vector-vector interaction on the general ground.  These are
all discussed in Sec.~\ref{sec:first}.

2) We will point out in Sec.~\ref{sec:three} that the shrinkage of the
first-order region in the $m_{ud}$-$m_s$ plane is not uncommon once we
take account of a vector-vector interaction.  We draw the critical
surface using the Polyakov loop augmented Nambu--Jona-Lasinio (PNJL)
model to exemplify the effect of the vector interaction, which may be
a likely explanation for the results by de~Forcrand and Philipsen.


\section{First-order Phase Transition}
\label{sec:first}

Let us first consider a simple chiral model at $T=0$ with and without
the vector interaction.  We will do so because we should understand
the underlying mechanism for possibility of the first-order phase
transition at $T=0$ and $\mu\neq0$ to elucidate the effect of the
vector interaction.

The simplest case in the chiral limit is enough for our present
demonstration in which one dynamical mass $M$ serves as the order
parameter.  Here $M$ is either the constituent quark mass in the color
deconfined phase or one third of the nucleon mass minus binding energy
of nuclear matter if the system confines color.  Although we can
formulate both, we shall focus on the former, i.e.\ deconfined quark
matter, hereafter.  (For the latter, see discussions in
Ref.~\cite{Halasz:1998qr}.)  We assume that the pressure $P_\chi[M]$
results in the chiral symmetry broken phase at zero density.  That is,
the free energy, $-P_\chi[M]$, has minima located at $M=\pm M_0$;  we
simply postulate the following form;
\begin{equation}
 P_\chi[M] = -a(M_0^2-M^2)^2
\label{eq:P_chi}
\end{equation}
with a parameter $a$.  Here we note that a linear term in $M$ should
be present if the current quark mass is nonzero.  We can neglect this
explicit chiral symmetry breaking in the qualitative level because
such a term has only minor effects on the phase transition in the
two-flavor sector.  In the three-flavor case, in contrast, the
$\mathrm{U_A}(1)$ breaking term generates a cubic term in $M$ which
favors the first-order phase transition.  We will not think of this
situation;  our purpose here is to see how the first-order transition
is possible at high density even though it is of second order at
vanishing density.  Thus, the above form of Eq.~(\ref{eq:P_chi}) is
valid when all the quarks are massless and the three-flavor
$\mathrm{U_A}(1)$ breaking is not significant.

Now let us turn finite $\mu$ on.  As long as $\mu$ is smaller than the
lowest-lying mass of fermionic excitation, nothing happens and the
vacuum remains empty.  Once $\mu$ exceeds a certain threshold $M$, a
finite amount of density appears.  The pressure has a contribution
from the density which is generally expressed as
\begin{equation}
 P_\mu[M] = \int_0^\mu d\mu' n_q(\mu') \,.
\label{eq:p_mu}
\end{equation}
Here $n_q(\mu)$ represents the fermion density.  In the quasi-particle
picture it is given by the integrated Dirac-Fermi distribution
function with the constituent mass;
\begin{align}
 n_q(\mu) &= \nu\int \frac{d^3p}{(2\pi)^3} \biggl[
  \frac{1}{e^{(\varepsilon-\mu)/T}+1} -
  \frac{1}{e^{(\varepsilon+\mu)/T}+1} \biggr] \notag\\
 &\overset{T=0}{\longrightarrow} \frac{\nu}{6\pi^2}(\mu^2-M^2)^{3/2}
  \,\theta(\mu^2-M^2) \,,
\label{eq:n}
\end{align}
where $\varepsilon=\sqrt{p^2+M^2}$ and $\nu$ is the fermionic degrees
of freedom (color$\times$flavor$\times$spin).  In two-flavor quark
matter, for relevant example,
$\nu=(3~\text{colors})\times(2~\text{flavors})\times(2~\text{spins})=12$.
We note that $\theta$ denotes the Heaviside theta function, which
signifies that the system at $\mu<M$ is empty.  In fact, the theta
function is essential to make a double-peak shape in the total
pressure, as we will see soon.

\begin{figure}
\includegraphics[width=0.42\textwidth]{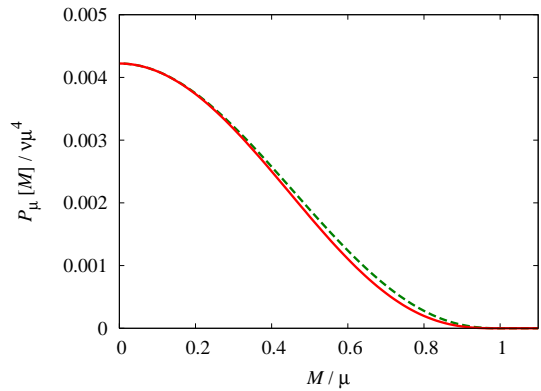}
\caption{Comparison between the exact integration in
  Eq.~(\ref{eq:p_mu}) (by the dotted curve) and the approximation in
  Eq.~(\ref{eq:approx}) (by the solid curve).}
\label{fig:approx}
\end{figure}

It is possible to perform the integration (\ref{eq:p_mu}) to find an
analytical expression with logarithmic terms.  To simplify our
qualitative analysis, however, we shall introduce an approximation as
\begin{equation}
 P_\mu[M] \approx \frac{\nu}{24\pi^2\mu^2}(\mu^2-M^2)^3\,
  \theta(\mu^2-M^2) \,,
\label{eq:approx}
\end{equation}
which turns out to be a good approximation as shown in
Fig.~\ref{fig:approx}.  The solid curve represents
Eq.~(\ref{eq:approx}), while the dotted curve is Eq.~(\ref{eq:p_mu})
with Eq.~(\ref{eq:n}) substituted.  Because more particles can reside
in the Fermi sphere for smaller mass, $P_\mu[M]$ has a maximum at
$M=0$.

\begin{figure}
\includegraphics[width=0.4\textwidth]{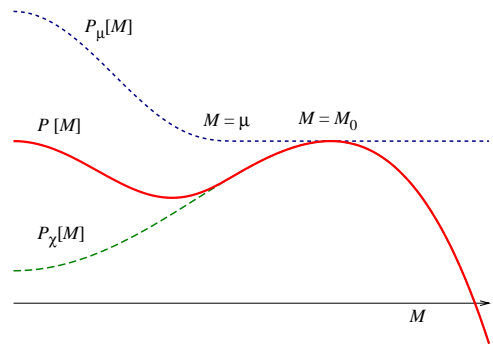}
\caption{Sketch of the double-peak pressure $P[M]$ resulting from the
  sum of $P_\chi[M]$ and $P_\mu[M]$.}
\label{fig:sketch}
\end{figure}

Let us consider the condition for $P[M]=P_\chi[M]+P_\mu[M]$ to have a
first-order phase transition.  Here $P_\chi[M]$ and $P_\mu[M]$ have a
peak at $M=M_0$ and $M=0$ respectively (see Fig.~\ref{fig:sketch}).
The existence of double peaks in $P[M]$ requires that
$\mu\lesssim M_0$, meaning that $\mu$ should not be much greater than
$M_0$.  [So, $\mu$ can be larger than $M_0$ slightly.]  This is
necessary for the peak at $M=M_0$ to survive.  At $M=0$ the pressure
curvature (i.e.\ the coefficient of the $M^2$ term) should be
negative, that is;
\begin{equation}
 a < \frac{\nu}{16\pi^2}\frac{\mu^2}{M_0^2} 
  \lesssim \frac{\nu}{16\pi^2} \,.
\label{eq:curvature}
\end{equation}
At the first-order critical point the peak at $M=0$ is as high as the
second peak at $M=M_0$ (neglecting a small shift by the contribution
from $P_\mu[M]$), which yields the critical condition that
\begin{equation}
 a \simeq \frac{\nu}{24\pi^2}\frac{\mu_c^4}{M_0^4} \,.
\label{eq:critical}
\end{equation}
As long as $\mu$ is raised with $\mu\lesssim M_0$ satisfied, the
curvature condition~(\ref{eq:curvature}) is sufficient for the
existence of $\mu_c$ deduced from Eq.~(\ref{eq:critical}).  This is
another way to see why we should have required $\mu\lesssim M_0$.

We shall next take account of the mean-fields from the vector-vector
interaction,
$-G_V(\bar{\psi}\gamma_\mu\psi)(\bar{\psi}\gamma^\mu\psi)$~\cite{Fukushima:2008wg,Kitazawa:2002bc,Sasaki:2006ws,Sakai:2008ga}.
Because $\langle\bar{\psi}\gamma^0\psi\rangle$ is nothing but the
fermionic density, roughly speaking, the vector interaction generates
a contribution to the pressure;
\begin{equation}
 P_V[M] = -G_V n_q^2 = -\frac{G_V \nu^2}{36\pi^4} (\mu^2-M^2)^3
  \,\theta(\mu^2-M^2)
\label{eq:vector}
\end{equation}
at $T=0$, which takes the same functional form as the approximated
(\ref{eq:approx}).  [In the above we have constructed $P_V[M]$ using
$n_q$ given by $\partial P_\mu/\partial\mu$.  This approximation is
qualitatively reasonable, but not self-consistent with the
\textit{full} pressure from which $n_q$ should have been inferred.  We
will come back to this point later.]  The coefficient in
Eq.~(\ref{eq:approx}) is hence modified by the effect of
Eq.~(\ref{eq:vector}) (that is, the effective fermionic degrees of
freedom are reduced), and the curvature condition is then
\begin{equation}
 a < \Bigl(1-\frac{2\nu G_V\mu^2}{3\pi^2}\Bigr)
  \frac{\nu}{16\pi^2}\frac{\mu^2}{M_0^2} \,.
\label{eq:modified}
\end{equation}
Therefore, even though we start with $a$ that satisfies
Eq.~(\ref{eq:curvature}), there is a critical $G_V$ for which
Eq.~(\ref{eq:modified}) is not satisfied within $\mu\lesssim M_0$.
Then, the first-order phase transition would disappear.

Let us see concrete numbers;  in the conventional NJL model with two
flavors~\cite{Hatsuda:1994pi}, for which our discussions based on
Eq.~(\ref{eq:P_chi}) are valid, one can read $a$ as
\begin{equation}
 a = \frac{1}{2M_0^2}\biggl( \frac{\nu\Lambda^2}{8\pi^2}
  -\frac{1}{4G_S} \biggr) = 0.067 \,,
\end{equation}
where we used $\nu=12$ and the ultraviolet cutoff $\Lambda=631\MeV$,
the four-fermion coupling $G_S\Lambda^2=2.19$, and the resultant
$M_0=336.2\MeV$~\cite{Hatsuda:1994pi}.  The first term originates from
the zero-point energy and the second from the four-fermion
interaction.  Then, the right-hand side of Eq.~(\ref{eq:curvature}) is
$0.076$ and the inequality is certainly satisfied.  In the case of the
chiral quark model~\cite{Gocksch:1991up}, for comparison, $a$ is
estimated as $a\simeq m_\sigma^2 f_\pi^2/(8M_0^4)$, which yields
$a=0.02\sim0.05$ depending on the value of the $\sigma$ meson mass.
This is again within the region of the first-order phase transition.

The critical point in the NJL model case is
\begin{equation}
 \mu_c=1.07M_0 \,,
\end{equation}
from Eq.~(\ref{eq:critical}).  This estimate is consistent with the
empirical value in the NJL model study.

It is easy to see the effect of $G_V$ from Eq.~(\ref{eq:modified}).
As we have confirmed the critical $\mu$ is nearly $M_0$ and we can
replace $\mu\to M_0$ in Eq.~(\ref{eq:modified}) approximately.  Then,
the inequality does not hold when
\begin{equation}
 G_V > 0.25 G_S \,,
\label{eq:critical_gv}
\end{equation}
meaning that such $G_V$ makes the first-order phase transition
disappear at all along the $\mu$ axis.  It is impressing that this
rough estimate is fairly consistent with a more serious analysis in
the NJL model~\cite{Kitazawa:2002bc}.

We can learn from the above argument that the repulsive vector-vector
interaction reduces the pressure arising from the degenerated
particles in the Fermi sphere.  To achieve chiral restoration with
such strong vector interaction, then, $\mu>M_0$ is necessary and a
peak around $M=M_0$ is  washed away by the tail in $P_\mu[M]+P_V[M]$
which extends up to $M=\mu$.  This is the qualitative mechanism how
the vector interaction would hinder the first-order phase transition.


\section{Three-Flavor Mass Plane}
\label{sec:three}
From the point of view of thermodynamic pressure, so far, we have seen
how the first-order phase transition could occur at
$\mu=\mu_c\simeq M_0$ especially in the case of two-flavor cold quark
matter.

Let us now proceed to the main part of our discussions on three-flavor
quark matter.  For a fixed value of $\mu$, the chiral phase
transition at finite $T$ is of first order when the current quark
masses $m_{ud}$ and $m_s$ stay small, and becomes of crossover if
$m_{ud}$ and $m_s$ exceed a critical boundary.  Therefore, in the 3D
space of $m_{ud}$, $m_s$, and $\mu$, the critical condition makes a
hypersurface.  For concreteness we adopt the PNJL model used in
Ref.~\cite{Fukushima:2008wg} to draw the critical surface.  The NJL
model might allow for artificial quark excitations at finite
temperature due to the lack of confinement.  This tends to shift the
location of the critical point $(\mu_E, T_E)$ to higher $\mu_E$ and
lower $T_E$ because the temperature effect smears the first-order
phase transition induced by density.  The PNJL model cures this (for
instance $T_E$ changes $\sim50\MeV$ in the NJL model to $\sim100\MeV$
in the PNJL model with the same parameter set~\cite{PNJL}) by means of
the color projection by the Polyakov loop which is a colored phase
factor associated with single-quark excitation.  Since the model
details are not our main focus, we simply refer to
Ref.~\cite{Fukushima:2008wg,PNJL} for details.

\begin{figure}
\includegraphics[width=0.47\textwidth]{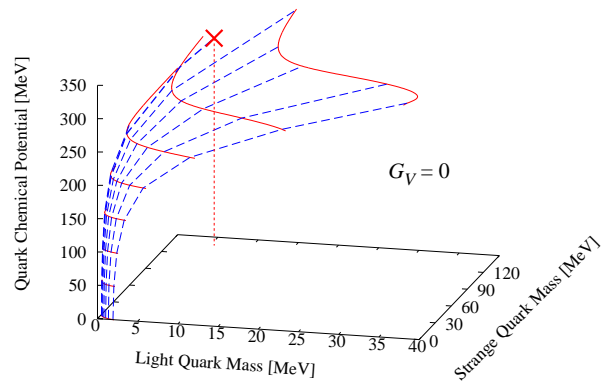}
\caption{Critical surface with standard curvature in the PNJL model
  without the vector interaction.  The physical mass point in this
  model is $(m_{ud}=5.5\MeV,\; m_s=135.7\MeV)$ which hits the critical
  surface at $\mu=\mu_E$.}
\label{fig:surface00}
\end{figure}

Figure~\ref{fig:surface00} is the critical surface with standard
curvature obtained in the PNJL model.  Above the critical surface the
finite-$T$ phase transition is of first order, while no sharp phase
transition takes place below the surface.  The second-order phase
transition sits on the critical surface.  This is just one model
example but shows typical behavior in model
studies~\cite{Kovacs:2006ym}.  The physical mass point hits the
critical surface at $\mu=\mu_E=313.5\MeV$ where the critical
temperature is $T=T_E=101.8\MeV$ (shown by a cross in
Fig.~\ref{fig:surface00}).

It is general in the quasi-particle picture that the density effect
induces a pressure like Eq.~(\ref{eq:p_mu}) whose maximum is located
in the (partially) chiral symmetric phase.  In the presence of such an
additional peak near $M_{ud}\simeq m_{ud}$, $M_s\simeq m_s$, the
first-order transition region is enhanced with increasing $\mu$.

The problem is that this curvature of the critical surface as shown in
Fig.~\ref{fig:surface00} might be inconsistent with the lattice
observation~\cite{deForcrand:2006pv} even though the model results
seem to be rather robust not relying on any special assumption.

\begin{figure}
\includegraphics[width=0.47\textwidth]{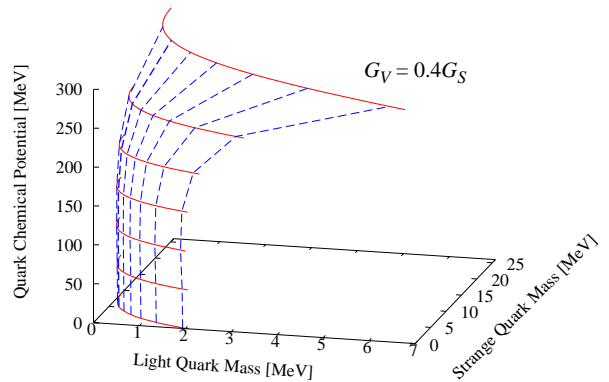}
\caption{Demonstration of the effect of the vector interaction in the
  PNJL model with $G_V=0.4G_S$.}
\label{fig:surface04}
\end{figure}

\begin{figure}
 \includegraphics[width=0.47\textwidth]{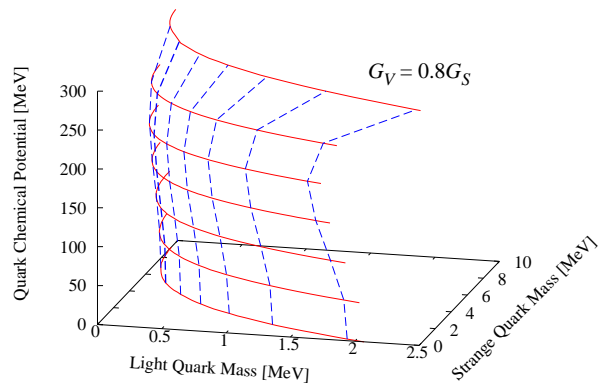}
\caption{Demonstration of the effect of the vector interaction in the
  PNJL model with $G_V=0.8G_S$.}
\label{fig:surface08}
\end{figure}

We here propose one scenario that has a natural account for the
lattice results and, at the same time, may not necessarily exclude the
existence of the critical point.  The necessary ingredient is the
vector-vector interaction alone.  As a matter of fact, the repulsive
vector interaction is anticipated from the hadron
property~\cite{Kitazawa:2002bc,Klimt:1990ws}.
Figures~\ref{fig:surface04} and \ref{fig:surface08} are the examples
from the PNJL model with the vector interaction incorporated.  Here we
should explain the mean-field treatment for the vector interaction in
a self-consistent
way~\cite{Asakawa:1989bq,Fukushima:2008wg,Kitazawa:2002bc,Sasaki:2006ws,Sakai:2008ga,Klimt:1990ws}.
The zeroth component in the vector interaction,
$-G_V(\bar{\psi}\gamma_0\psi)(\bar{\psi}\gamma^0\psi)$, produces the
mean-field terms, $-2G_V n_q\bar{\psi}\gamma^0\psi+G_V n_q^2$.  The
former term adds to the chemical potential as $\mu\to\mu-2G_V n_q$ and
the latter adjusts the larger contribution from the former.  Then, the
mean-field $n_q$ is fixed by $\partial P/\partial n_q=0$ which
guarantees the thermodynamic relation;
$n_q-\partial P/\partial\mu=0$.  [Note that
$\partial P/\partial n_q=-2G_V\partial P/\partial\mu$.]

We have chosen a considerably large value of $G_V$ to make it easier
to perceive the effect visually.  The point is that the inclusion of
$G_V\neq0$ induces the opposite curvature to that in
Fig.~\ref{fig:surface00} at small $\mu$, and eventually the curvature
returns to the standard one at larger $\mu$.  It should be noted that
the first-order region at $\mu=0$ does not change but the axis scale
in Figs.~\ref{fig:surface04} and \ref{fig:surface08} is different from
Fig.~\ref{fig:surface00}.

\begin{figure}
 \includegraphics[width=0.42\textwidth]{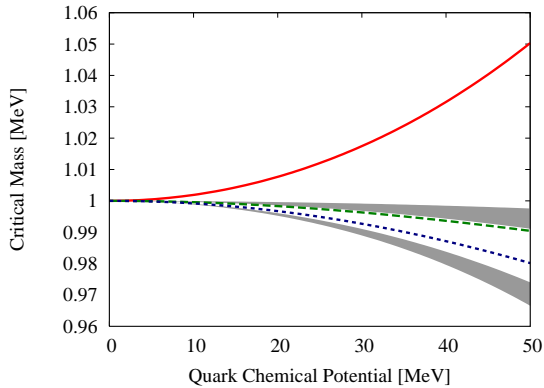}
\caption{Critical mass $m_c$ along the $m_{ud}=m_s$ line as a function
of $\mu$.  The (red) solid curve represents the PNJL model results at
$G_V=0$, the (green) dashed curve at $G_V=0.4G_S$, and the (blue)
dotted curve at $G_V=0.8G_S$.  The upper and lower shaded regions
correspond to the lattice results reported in the first and last
papers in Ref.~\cite{deForcrand:2006pv}, respectively.}
\label{fig:curvature}
\end{figure}

Although our aim is just to present some demonstrations like
Figs.~\ref{fig:surface04} and \ref{fig:surface08}, it is intriguing to
make a quantitative comparison.  Along the diagonal $m=m_{ud}=m_s$
line, the critical mass is expanded in terms of $\mu$ as
\begin{equation}
 \frac{m_c(\mu)}{m_c(0)} = 1
  +c_2\biggl(\frac{\mu}{\pi T_c}\biggr)^2
  +c_4\biggl(\frac{\mu}{\pi T_c}\biggr)^4 +\cdots\,.
\end{equation}
Our calculations give $c_2=5.88$ and $c_4=43.8$ at $G_V=0$ as shown by
the solid curve in Fig.~\ref{fig:curvature}.  The lattice results are,
on the other hand, $c_2=-0.7(4)$ and $c_4\simeq0$ in the first paper
of Ref.~\cite{deForcrand:2006pv} which is indicated by the upper
shaded region in Fig.~\ref{fig:curvature}.  In the last paper of
Ref.~\ref{fig:curvature} significantly different values are reported;
$c_2=-3.3(3)$ and $c_4=-47(20)$ as shown by the lower shaded region.
In our calculations the vector interaction drastically alters the
curvature from the solid curve to the dashed (dotted) curve in
Fig.~\ref{fig:curvature} at $G_V=0.4G_S$ ($G_V=0.8G_S$).

From Fig.~\ref{fig:curvature} we may say that some $G_V<0.4G_S$ could
be enough to understand the results in the first paper of
Ref.~\cite{deForcrand:2006pv}, while the vector interaction alone is
not sufficient to reproduce the results in the last paper of
Ref.~\cite{deForcrand:2006pv}.  To make the statement more conclusive
we need more lattice QCD data.

The rest of this paper will be devoted to qualitative explanation of
this back-bending curvature as a result of the vector interaction.  We
will discuss the mechanism in order.

1) We have to realize that the finite $T$ is important to understand
the back-bending behavior.  The information of the critical
temperature is implicit in Figs.~\ref{fig:surface00},
\ref{fig:surface04}, and \ref{fig:surface08}.  The general tendency is
that the critical temperature becomes smaller as $\mu$ increases.
Thus, in fact, the critical surface is cut at large chemical potential
for which the critical point touches $T=0$.

\begin{figure}
\includegraphics[width=0.42\textwidth]{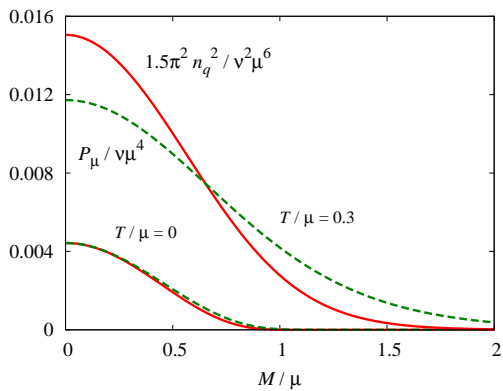}
\caption{$n_q^2$ (solid curve) and $P_\mu$ (dotted curve) at $T/\mu=0$
  and $T/\mu=0.3$, where $n_q^2$ is normalized such that $P_\mu$ and
  $n_q^2$ nearly coincide at $T/\mu=0$.}
\label{fig:T}
\end{figure}

2) In the vicinity of $\mu=0$, hence, the finite $T$ effect is
substantial, which makes the functional forms of the integrated
$n_q(\mu)$ with respect to $\mu$ (i.e.\ $P_\mu$) and $n_q^2$ deviate
from each other unlike the $T=0$ case.  It should be mentioned that
the simple argument in Sec.~\ref{sec:first} is still applicable for
$m_{ud}=m_s$ as it is (except for the $\mathrm{U_A}(1)$ breaking
term).  Then the integrated one~(\ref{eq:p_mu}) turns out to be less
steeper than $n_q^2$ as a function of $M$, which can be readily
confirmed by simple numerical calculations (see Fig.~\ref{fig:T}).

Thus the pressure contribution by density-induced particles is
relatively more suppressed by finite $T$ than the vector interaction
effect involving $n_q^2$.  This means, in other words, that $n_q^2$
brings a sharper modification than $P_\mu$ to the total pressure at
higher $T$ (and thus smaller $\mu$ in turn).

3) Because the repulsive vector interaction has the opposite effect to
the density, as we have seen in Sec.~\ref{sec:first}, some $G_V$
exists which is large enough to invert the density effect.  Still, the
first-order region at $\mu=0$ is intact since $n_q=0$ at $\mu=0$.

4) With appropriate $G_V\neq0$, as $\mu$ goes up, the first-order
region shrinks by the effect of growing $n_q^2$ from the vector
interaction which overwhelms the effect of $P_\mu$ as long as $\mu/T$
is small.

5) When $\mu$ gets larger and $T/\mu$ becomes smaller, the functional
shape of $n_q^2$ comes to be identical to that of $P_\mu$ as seen in
Fig.~\ref{fig:T}.  If $G_V$ is not too large, the density effect can
surpass the vector interaction eventually, which makes the curvature
bend back into the standard direction at high $\mu$, which is
manifestly the case in Figs.~\ref{fig:surface04} and
\ref{fig:surface08}.


\section{Summary}
We have clarified how the quasi-particle description can lead to a
first-order phase transition in cold and dense quark matter.  In the
same way we have intuitively made clear the role played by the
mean-fields from the repulsive vector-vector interaction which reduces
a pressure contribution from density-induced particles.

We have then discussed the vector interaction in the $m_{ud}$-$m_s$
plane with three flavors.  We have drawn the critical surface using
the PNJL model with the vector interaction.  Along the same line as
the quasi-particle description we have given a simple account for the
back-bending behavior of the critical surface, which is consistent
with the negative curvature in the recent lattice simulations.
Logically speaking, therefore, there might be a chance that the
critical point still persists even with the negative curvature at
small density.

It should be remarked, however, that whether the back bending occurs
or not cannot be naively interpreted as whether the critical point
exists or not.  This is because, as we mentioned, the critical surface
is cut at some large $\mu$ where the critical temperature is zero.  As
a matter of fact, the critical value of $G_V$ is around $0.25$ as in
Eq.~(\ref{eq:critical_gv}) (see also
Refs.~\cite{Fukushima:2008wg,Kitazawa:2002bc,Sasaki:2006ws}) for which
we cannot observe clear back-bending behavior in the PNJL model.  In
this way, if we take the comparison to the PNJL model analysis
seriously, the lattice results by de~Forcrand and Philipsen certainly
suggest the nonexistence of the QCD critical point.

There are important issues to be investigated in the future.  First of
all, the determination of $G_V$ is indispensable.  For this purpose,
unfortunately, the description of the vector meson property within the
NJL model is not quite reliable because the momentum cutoff is not
large enough as compared to the vector meson mass.  The lattice
simulation is one possibility, though observable to measure $G_V$ is
not clear yet.  Second, not only the vector interaction but also the
anomaly ('t~Hooft) term is important for the existence of the QCD
critical point~\cite{Pisarski:1983ms}.  The density dependence of the
anomaly coupling strength is unknown from the lattice simulation,
which may affect the critical surface curvature.  The density
dependence of $G_V$ is not known, either, but the point of what we
have shown here is that intriguing density dependence appears from
$n_q^2$ even for a constant $G_V$.  Hence, if the opposite curvature
of the critical surface to the standard one is really established in
the lattice simulation, it could be interpreted as a circumstantial
evidence for finite repulsive vector-vector interaction.


\acknowledgments
The author is grateful to Teiji Kunihiro and Akira Ohnishi for
discussions.  He thanks the Institute for Nuclear Theory at the
University of Washington for its hospitality and the Department of
Energy for partial support during the completion of this work.  There,
he benefited from discussions with Misha Stephanov, Volker Koch,
Philippe de~Forcrand, Sourendu Gupta, and Maria-Paola Lombardo.  He is
supported by Japanese MEXT grant No.\ 20740134 and also supported in
part by Yukawa International Program for Quark Hadron Sciences.


\end{document}